\documentclass[twocolumn,showpacs,preprintnumbers,amsmath,amssymb,times]{revtex4}
\usepackage{epsfig}

\usepackage{graphicx} 
\usepackage{url}      
\usepackage{amsmath}  

\begin{document}

\title{Traffic dynamics of packets generated with non-homogeneously selected sources and destinations
in scale-free networks}

\author{Rui Jiang${}^1$, Mao-Bin Hu${}^1$, Wen-Xu Wang${}^2$, Gang Yan${}^3$, Qing-Song Wu${}^1$, Bing-Hong Wang${}^2$ }
\address{${}^1$ School of Engineering Science, University of Science and technology of China, Hefei 230026, China}
\address{${}^2$ Nonlinear Science Center and Department of Modern Physics,
University of Science and technology of China, Hefei 230026, China}
\address{${}^3$ Department of Electronic Science and Technology,
University of Science and technology of China, Hefei 230026, China}


\begin{abstract}
In this paper, we study traffic dynamics in scale-free networks in
which packets are generated with non-homogeneously selected
sources and destinations, and forwarded based on the local routing
strategy. We consider two situations of packet generation: (i)
packets are more likely generated at high-degree nodes; (ii)
packets are more likely generated at low-degree nodes. Similarly,
we consider two situations of packet destination: (a) packets are
more likely to go to high-degree nodes; (b) packets are more
likely to go to low-degree nodes. Our simulations show that the
network capacity and the optimal value of $\alpha$ corresponding
to the maximum network capacity
greatly depend on the configuration of packets' sources and
destinations. In particular, the capacity is greatly enhanced when
most packets travel from low-degree nodes to high-degree nodes.

\end{abstract}
\maketitle

\section{Introduction}

Complex networks can describe a wide range of systems in nature and
society, therefore there has been a quickly growing interest in this
area [1-3]. Since the surprising small-world phenomenon discovered
by Watts and Strogatz [4] and scale-free phenomenon with degree
distribution following $P(k)\sim k^{-\gamma}$ by Barab\'{a}si and
Albert[5], the evolution mechanism of the structure and the dynamics
on the networks have recently received a lot of interests among
physics community. Due to the importance of large communication
networks such as the Internet and WWW with scale-free properties in
modern society, processes of dynamics taking place upon the
underlying structure such as traffic congestion of information flow
have drawn more and more attention from physical and engineering
fields.

The ultimate goal of studying these large communication networks
is to control the increasing traffic congestion and improve the
efficiency of information transportation. Many recent studies have
focused on the efficiency improvement of communication networks
which is usually considered from two aspects: modifying underlying
network structures or developing better routing strategies. In
view of the high cost of changing the underlying structure, the
latter is comparatively preferable.

Recent works proposed some models to mimic the traffic routing on
complex networks by introducing packets generating rate as well as
homogeneously selected sources and destinations of each packet
[6-12]. These kinds of models also define the capacity of networks
measured by a critical generating rate. At this critical rate, a
continuous phase transition from free flow state to congested
state occurs. In the free state, the numbers of created and
delivered packets are balanced, leading to a steady state. While
in the jammed state, the number of accumulated packets increases
with time due to the limited delivering capacity or finite queue
length of each node. In these models, packets are forwarded
following the random walking [6], the shortest path [7], the
efficient path [8], the next-nearest-neighbor search strategy [9],
the local information [10] or the integration of local static and
dynamic information [11,12].

Nevertheless, in previous studies, packets are generated with
homogeneously selected sources and destinations, i.e., sources and
destinations are randomly chosen without preference. However, in
the real networked traffic, packets are more likely to be
generated at some nodes than at others and are more likely to go
to some nodes than to others. Therefore, in this paper, we study
traffic dynamics with considering packets are generated with
non-homogeneously selected sources and destinations, and delivered
based on the local routing strategy, which is favored in cases
where there is a heavy communication cost to searching the
network.

The paper is organized as follows. In section 2, the traffic model
is introduced. In section 3, the simulations results are presented
and discussed. The conclusion is given in section 4.

\section{Model and rules}

Barab\'{a}si-Albert model is the simplest and a well known model
which can generate networks with power- law degree distribution
$P(k)\sim k^{-\gamma}$, where $\gamma = 3$. Without losing
generality, we construct the network structure by following the
same method used in Ref. [5]: Starting from $m_0$ fully connected
nodes, a new node with $m_0$ edges is added to the existing graph
at each time step according to preferential attachment, i.e., the
probability $\prod_i$ of being connected to the existing node $i$
is proportional to the degree $k_i$.

Then we model the traffic of packets on the given graph. At each
time step, there are $R$ packets generated in the system. Each
packet is generated at node $i$ with probability $
\frac{k_i^{\beta_1}}{\sum k^{\beta_1}}, $ where $k_i$ is the
degree of node $i$ and the sum runs over all nodes. Furthermore,
the packet goes to the node $j$ with probability $
\frac{k_j^{\beta_2}}{\sum k^{\beta_2}}, $ where the sum also runs
over all nodes. Here $\beta_1$ and $\beta_2$ are two parameters.
In the special case of $\beta_1=\beta_2=0$, the packets are
generated with homogeneously selected sources and destinations.
When $\beta_1>0$ ($\beta_1<0$), packets are more likely generated
at high-degree (low-degree) nodes. When $\beta_2>0$ ($\beta_2<0$),
packets are more likely to go to high-degree (low-degree) nodes.

We treat all the nodes as both hosts and routers and assume that
each node can deliver at most $C$ packets per time step towards
their destinations. All the nodes perform a parallel local search
among their immediate neighbors. If a packet's destination is found
within the searched area of node $l$, i.e. the immediate neighbors
of $l$, the packets will be delivered from $l$ directly to its
target and then removed from the system. Otherwise, the probability
of a neighbor node $i$, to which the packet will be delivered is as
follows:
\begin{equation}
P_{l\rightarrow i} = \frac{k_i^{\alpha}}{\sum_j k_j^{\alpha}} ,
\end{equation}
where the sum runs over the immediate neighbors of the node $l$.
$\alpha$ is an introduced tunable parameter. During the evolution
of the system, FIFO (first- in first-out) rule is applied.  All
the simulations are performed with choosing $C = 10$.

In order to characterize the network capacity, we use the order
parameter presented in Ref. [13]:
\begin{equation}
\eta(R) = \lim_{ t\rightarrow \infty} \frac{<\triangle Load>}{R
\triangle t},
\end{equation}
where $Load(t)$ is defined as the number of packets within the
network at time $t$. $\triangle Load = Load(t+\triangle
t)-Load(t)$ with $<\cdots>$ indicates average over time windows of
width $\triangle t$. The order parameter represents the ratio
between the outflow and the inflow of packets calculated over long
enough period. In the free flow state, due to the balance of
created and removed packets, the load is independent of time,
which brings a steady state. Thus when time tends to be unlimited,
$\eta$ is about zero. Otherwise, when $R$ exceeds a critical value
$R_c$, the packets will continuously pile up within the network,
which destroys the stead state. Thus, the quantities of packets
within the system will be a function of time, which makes $\eta$ a
constant more than zero. Therefore, a sudden increment of $\eta$
from zero to nonzero characterizes the onset of the phase
transition from free flow state to congestion, and the network
capacity can be measured by the maximal generating rate $R_c$ at
the phase transition point.

\section{Results}


In the special case of $\beta_1=\beta_2=0$ (case 1),  the maximum
network capacity is $R_c \approx 40$, which is reached at the
optimal value $\alpha_c \approx -1$. This can be explained by
noting two facts that the degree-degree correlation is zero in BA
networks and the average number of packets on nodes does not
depend on degree $k$ when $\alpha_c \approx -1$ [10].


Next we investigate the case of $\beta_1=-5.0$ and $\beta_2=-5.0$
(case 2), i.e., most packets travel from low-degree nodes to
low-degree nodes. Fig.1 compares the network capacity $R_c$ in
cases 1 and 2. One can see that at a given $\alpha$, the network
capacity decreases. This is easily to be understood because a
low-degree node has less links and therefore more difficult to be
found by packets than a high-degree node.

Furthermore, the optimal value $\alpha_c$ is essentially the same
in cases 1 and 2, which is explained as follows. Let $n_i$ denote
the number of packets of node $i$ at time $t$. Then we have
\begin{equation}
\frac{dn_i}{dt}=-n_{\mbox{deliver}}+n_{\mbox{receive}}+n_{\mbox{generate}}-n_{\mbox{remove}},
\end{equation}
where $n_{\mbox{deliver}}, n_{\mbox{receive}},
n_{\mbox{generate}}$, and  $n_{\mbox{remove}}$ denote the number
of packets delivered from node $i$ to other nodes, received from
other nodes, generated at node $i$ and removed at node $i$. In
case 1, $n_{\mbox{generate}}=n_{\mbox{remove}}$, thus
\begin{equation}
\frac{dn_i}{dt}=-n_{\mbox{deliver}}+n_{\mbox{receive}}.
\end{equation}
From Eq.(4), Wang et al. show that $n(k)\sim k^{1+\alpha}$ [10].
Therefore, when $\alpha=-1$, the average number of packets on
nodes is independent of degree $k$ and the maximum capacity is
reached.

In case 2, we have $n_{\mbox{generate}}=n_{\mbox{remove}}>0$ for
low-degree nodes and $n_{\mbox{generate}}=n_{\mbox{remove}}
\approx 0$ for high-degree nodes. Thus, Eq.(4) is valid for both
low-degree nodes and high-degree nodes. Therefore, the optimal
value $\alpha_c \approx -1$ does not change.

\begin{figure}
\centering \epsfig{file=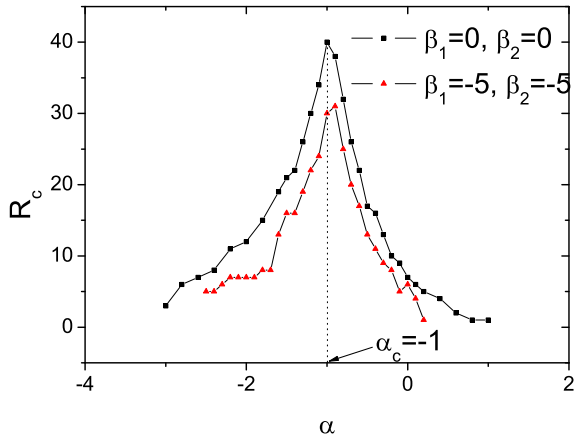, width=8cm} \caption{(color online)
The network capacity $R_c$ against $\alpha$ in cases 1 and 2. }
\end{figure}






Fig.2 compares the network capacity $R_c$ in case 1 and in case 3
(where $\beta_1=-2.0$ and $\beta_2=2.0$) and case 4 (where
$\beta_1=-5.0$ and $\beta_2=5.0$), i.e., most packets travel from
low-degree nodes to high-degree nodes. One can see that the
network capacity is greatly enhanced in cases 3 and 4. The maximum
network capacity increases from 40 in case 1 to 119 in case 3 and
to 720 in case 4. This is because a high-degree node has much more
links and therefore much easier to be found by packets than a
low-degree node. Based on this result, we suggest that the local
routing strategy is very suitable if the packets are more likely
to go from low-degree nodes to high-degree nodes.

Moreover, the optimal value of $\alpha_c$ corresponding to the
maximum capacity increases from $-1.0$ in case 1 to $-0.9$ in case
3 and to $-0.8$ in case 4. This is also explained from Eq.(3). In
case 1, $\alpha_c>-1$ means high-degree nodes have more packets.
In cases 3 and 4, we have $n_{\mbox{generate}}>0,
n_{\mbox{remove}}\approx 0$ for low-degree nodes and
$n_{\mbox{generate}}\approx 0, n_{\mbox{remove}}> 0$ for
high-degree nodes. The packets generated in low-degree nodes and
removed in high-degree nodes enable the average number of packets
on nodes to be independent of degree $k$. As a result, the maximum
capacity is achieved.

\begin{figure}
\centering \epsfig{file=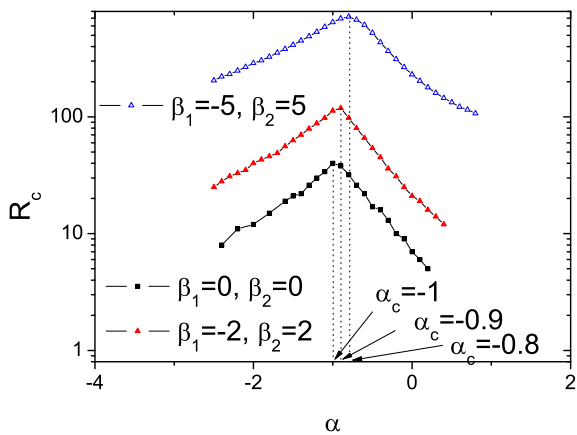, width=8cm} \caption{(color online)
The network capacity $R_c$ against $\alpha$ in cases 1, 3 and 4. }
\end{figure}


Fig.3 compares the network capacity $R_c$ in case 1 and in case 5,
where $\beta_1=5.0$ and $\beta_2=-5.0$, i.e., most packets travel
from high-degree nodes to low-degree nodes. One can see that the
network capacity becomes smaller and the optimal value of
$\alpha_c$ decreases. The reason of capacity decrease is the same
as in case 2, i.e., a low-degree node has less links and therefore
more difficult to be found by packets. The decrease of $\alpha_c$
is explained as follows. In case 1, $\alpha_c<-1$ means
high-degree nodes have less packets. In case 5, we have
$n_{\mbox{generate}}>0, n_{\mbox{remove}}\approx 0$ for
high-degree nodes and $n_{\mbox{generate}}\approx 0,
n_{\mbox{remove}}> 0$ for low-degree nodes. The packets generated
in high-degree nodes and removed in low-degree nodes enable the
average number of packets on nodes to be independent of degree
$k$. Consequently, the maximum capacity is reached at
$\alpha_c<-1$.

\begin{figure}
\centering \epsfig{file=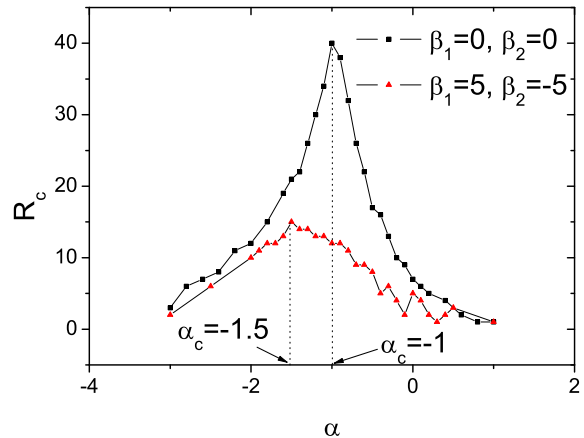, width=8cm} \caption{(color online)
The network capacity $R_c$ against $\alpha$ in cases 1 and 5. }
\end{figure}


Fig.4 compares the network capacity $R_c$ in case 1 and in case 6
(where $\beta_1=3.0$ and $\beta_2=3.0$) and case 7 (where
$\beta_1=5.0$ and $\beta_2=5.0$), i.e., most packets travel from
high-degree nodes to high-degree nodes. In case 6, the network
capacity is essentially independent of $\alpha$ for $\alpha<-1$
and decreases with the increase of $\alpha$. In case 7, the
network capacity is essentially independent of $\alpha$ when
$\alpha$ is in the range studied. This is explained as follows.

In case 6, the probability that the highest-degree node is chosen
as origin is 0.33 and it is 0.6 in case 7. Therefore, in case 6,
when $R_c>31$, the number of particles generated in the
highest-degree node exceeds the capacity of the node. This leads
to the congestion. As a result, the constant network capacity for
$\alpha<-1$ occurs. Similarly, in case 7, when $R_c>17$, the
number of particles generated in the node exceeds the capacity of
the node. Therefore, the constant network capacity in the studied
range emerges. To avoid the constant small network capacity, it is
necessary to enhance the capacity of the nodes of high degrees.

\begin{figure}
\centering \epsfig{file=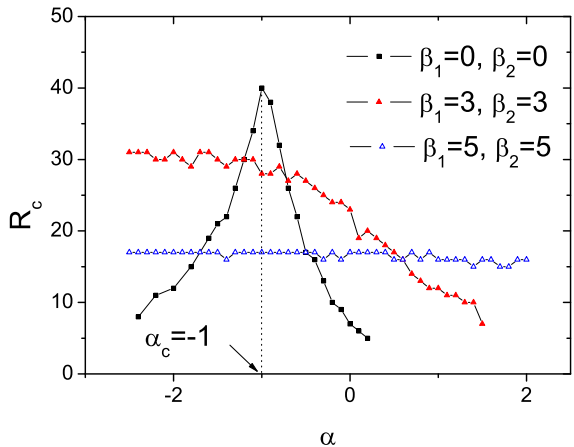, width=8cm} \caption{(color online)
The network capacity $R_c$ against $\alpha$ in cases 1, 6 and 7. }
\end{figure}

\section{Discussion and conclusion}

In this paper, we have investigated the network capacity in the
scale-free networks, in which packets are generated with
non-homogeneously selected sources and destinations, based on the
local routing strategy.

Generally speaking, when most packets travel to low-degree nodes,
the network capacity will decrease. In contrast, when most packets
travel to high-degree nodes, whether the network capacity
decreases or increases depends on the selection of origins. When
$\beta_2$ is large, i.e., most packets are generated from
high-degree nodes, the highest-degree node is easily congested,
which leads to the congestion of the whole network. To avoid this,
it is necessary to enhance the capacity of high-degree nodes. When
most packets are generated from low-degree nodes, the network is
greatly enhanced. Therefore, the local routing strategy is very
suitable if the packets are more likely to go from low-degree
nodes to high-degree nodes. 

In addition, $\alpha_c$, i.e., the optimal value of $\alpha$
corresponding to the maximum network capacity also depends on the
distribution of packets' origins and destinations. We have
explained the reason why $\alpha_c$ changes when the distribution
of packets' origins and destinations changes.

Finally, we would like to mention that our results may be used to
design a new local routing strategy, in which the parameter
$\alpha$ is packet-related. Concretely, $\alpha$ depends on the
origin and destination of the packet $\alpha=\alpha(k_o,k_d)$,
where $k_o$ and $k_d$ denote the degree of the node where the
packet is generated and that of the node where the packet goes to.
A suitable choice of $\alpha(k_o,k_d)$ may enhance the network
capacity. Further investigations will be carried out in future
work.

\section*{Acknowledgements}
We acknowledge the support of National Basic Research Program of
China (2006CB705500), the National Natural Science Foundation of
China (NNSFC) under Key Project No. 10532060 and Project Nos.
10404025, 10672160, 70601026, and the CAS special Foundation.


\end{document}